\documentclass{aastex701}
\usepackage{amsmath}
\usepackage{placeins}
\begin{document}
\title{How Intensity Threshold Shapes the Velocity Dispersion--Size Relation in Molecular Clouds}

\correspondingauthor{Fujun Du}
\email{fjdu@pmo.ac.cn}

\author[0009-0002-4323-7203]{Yaru Qi}
\affiliation{Purple Mountain Observatory, Chinese Academy of Sciences, Nanjing 210023, People's Republic of China}
\affiliation{School of Astronomy and Space Science, University of Science and Technology of China, Hefei 230026, People's Republic of China}
\email{yrqi@pmo.ac.cn}

\author[0000-0002-0409-7466]{Xin Liu}
\affiliation{Purple Mountain Observatory, Chinese Academy of Sciences, Nanjing 210023, People's Republic of China}
\email{liuxin@pmo.ac.cn}

\author[0000-0002-7489-0179]{Fujun Du}
\affiliation{Purple Mountain Observatory, Chinese Academy of Sciences, Nanjing 210023, People's Republic of China}
\email{fjdu@pmo.ac.cn}

\begin{abstract}
The velocity dispersion--size relation of molecular clouds, $\sigma_v = C \cdot (R/1\,\mathrm{pc})^\gamma$, has been reported with a power-law index ranging from $\gamma \approx 0.38$ to $0.5$ or higher, and a correspondingly varying normalization coefficient $C$, but the role of the intensity threshold in producing this scatter has not been quantitatively characterized. Using 4916 $^{13}$CO clouds identified with GDCluster \citep{liu2025} in the Milky Way Imaging Scroll Painting (MWISP) survey, we derive $\gamma$ and $C$ as functions of the intensity threshold $T_\mathrm{th}$. The index $\gamma$ increases monotonically with threshold, reaching a saturation value $\gamma_\mathrm{sat} \approx 0.56$ for the full sample; splitting the sample by spiral arm gives higher values, $\gamma_\mathrm{sat} \approx 0.62$ and $0.70$. The normalization $C(T_\mathrm{th})$ declines monotonically, saturating at $C_\mathrm{sat} \approx 0.35$~km\,s$^{-1}$ and implying a surface density of $\sim$46~$M_\odot\,\mathrm{pc}^{-2}$ under a virial approximation. On the $\gamma(T_\mathrm{th})$ curve, $\gamma \approx 0.38$ \citep{larson1981} and $\gamma \approx 0.5$ \citep{solomon1987} correspond to a low threshold ($T_\mathrm{th} \approx 1.2$~K) and a near-saturation one ($2.6$~K), respectively, roughly consistent with the $^{13}$CO-equivalent boundary temperatures of these two studies. These two historically debated values, together with the broader $\gamma \approx 0.38$--$0.7$ range reported in the literature, may largely reflect different positions along this single threshold-dependent curve and differences in sample composition. This threshold effect should therefore be considered when using Larson-type relations to assess the roles of turbulence and gravity. As an independent check, DBSCAN-identified $^{13}$CO and $^{12}$CO samples show the same monotonic trend in $\gamma$, though the curve shape differs quantitatively.
\end{abstract}

\keywords{\uat{Molecular clouds}{1072} --- \uat{Scaling relations}{2031} --- \uat{Interstellar dynamics}{839} --- \uat{CO line emission}{262}}

\section{Introduction} \label{sec:intro}

\citet{larson1981} compiled molecular cloud data from the literature and found a power-law relation, now known as Larson's first law, between velocity dispersion $\sigma_v$ and size $R$, $\sigma_v \propto R^\gamma$, with an index of $\gamma \approx 0.38$. \citet{solomon1987} redefined cloud boundaries for 273 clouds as closed isointensity surfaces of antenna temperature in position--position--velocity space, adopting $T_\mathrm{min} \approx 4$~K for most of the sample, and obtained $\gamma \approx 0.5$, consistent with clouds in virial equilibrium at roughly constant surface density. These two relations continue to be discussed and cited in the molecular cloud literature \citep{elmegreen2004, mckee2007, heyerdame2015}.

However, the $\sigma_v$--$R$ relation does not converge on a single universal form \citep{falgarone1992}: both its normalization and its power-law index have been reported to vary across studies. \citet{heyerbrunt2004} found that the structure functions of individual molecular clouds are described by a narrow range of scaling exponents and coefficients, consistent with a nearly universal turbulent structure across a wide range of cloud environments. A later reexamination of the \citet{solomon1987} sample using high-quality $^{13}$CO data \citep{heyer2009}, however, found that the normalization coefficient $v_0 = \sigma_v/R^{0.5}$ varies systematically with surface density $\Sigma$ rather than remaining constant, revising this earlier picture and showing that the $\sigma_v$--$R$ relation is less universal than \citet{larson1981} implied. The power-law index $\gamma$ is similarly unsettled. Subsequent studies have reported intermediate values: \citet{romanduval2010} obtained $\gamma \approx 0.39$ for 580 $^{13}$CO molecular clouds from the Galactic Ring Survey; \citet{rice2016} found $\gamma \approx 0.43$ using a dendrogram algorithm applied to the \citet{dame2001} survey; \citet{feng2024} obtained $\gamma \approx 0.43$ from Milky Way Imaging Scroll Painting (MWISP) $^{13}$CO structures combined with Gaia distances; and \citet{jiang2025} obtained $\gamma \approx 0.396$ from 71,000 MWISP $^{13}$CO clumps.

At the high end of the reported range, a number of studies have obtained steeper indices of $\gamma \gtrsim 0.6$ \citep[e.g.,][]{shetty2012, hughes2013, kauffmann2017, liu2021, zhou2022}. \citet{miville2017} used a full Galactic plane $^{12}$CO survey to build a catalog of 8107 clouds and obtained $\gamma \approx 0.63$. \citet{sun2024} used MWISP $^{12}$CO data to build a catalog of over 32,000 outer-Galaxy molecular clouds and found $\gamma \approx 0.66$. On the theoretical and simulation side, \citet{xie2025} proposed a two-component model combining shear and self-gravity, finding that the index for large-scale clouds can vary between 0.5 and 1 depending on environment; \citet{colman2024} found in simulations that the index for large clouds can reach 0.75, and identified cloud extraction algorithms as an important source of inconsistency in reported scaling relations. These results show that the scatter in $\gamma$ has multiple origins.

\citeauthor{larson1981}'s (\citeyear{larson1981}) sample came from a compilation of cloud catalogs with varied boundary definitions---some set by observational sensitivity, others by closed contours. It has been recognized that such boundary choices introduce systematic biases into the derived scaling relations \citep{kegel1989,bpml2002,rosolowsky2006,lombardi2010}. \citet{solomon1987} adopted a higher, more consistent isointensity surface threshold. These different threshold choices directly affect the measured $\gamma$, but the full variation of $\gamma$ as the threshold varies continuously from zero to high values has never been directly measured.

This paper quantifies how the power-law index $\gamma$ and the normalization coefficient $C$ of the $\sigma_v$--$R$ relation change as the intensity threshold is varied continuously from zero to high values. Using $^{13}$CO data from the MWISP survey, we recompute the physical parameters (size, velocity dispersion, etc.) of a sample of molecular clouds at a series of intensity thresholds, fit the Larson relation for each threshold, and build a quantitative description of the Larson parameters ($\gamma$ and $C$) as functions of the threshold. We aim to use this framework to understand the scatter in the reported $\gamma$ values in the literature.

\section{Data and Method} \label{sec:data}

We use $^{13}$CO ($J=1$--$0$) data from the MWISP survey, with an angular resolution of $\sim52''$ and a velocity resolution of $\sim$0.17~km\,s$^{-1}$ \citep{su2019, yang2026}. We use $^{13}$CO rather than $^{12}$CO for two reasons: first, $^{12}$CO emission is more extended and traces predominantly diffuse gas, whereas $^{13}$CO traces less diffuse and moderately compact structures; second, $^{12}$CO suffers from more severe line saturation, making cloud decomposition less reliable and causing adjacent clouds to blend into large continuous structures at low intensity thresholds. Clouds are identified using GDCluster \citep{liu2025}, an algorithm that incorporates spatial-continuity constraints during both Gaussian decomposition and subsequent clustering to separate kinematically blended velocity components, thereby reducing structure misidentification in crowded regions of the Galactic plane. GDCluster is applied to the MWISP $^{13}$CO ($J=1$--$0$) survey data covering the second Galactic quadrant ($104\fdg75 \leq l \leq 150\fdg25$, $-5\fdg25 \leq b \leq 5\fdg25$). Distances are determined exclusively via the kinematic method, with $v_\mathrm{LSR}$ converted to distance using the rotation curve model of \citet{reid2019}. To ensure reliable kinematic distances, we exclude clouds with $|v_{\rm LSR}| < 10$~km\,s$^{-1}$. This leaves a final sample of 4916 molecular clouds. For $|v_{\rm LSR}|$ below 10~km\,s$^{-1}$, the errors in kinematic distance become very large \citep{yan2020}, consistent with the $\lesssim 10$~km\,s$^{-1}$ non-circular motions typical of spiral-arm tracers \citep{reid2019}. Sensitivity to this $|v_{\rm LSR}|$ cut is tested in Appendix~\ref{subsec:app_vcut}. The effective radius to be used in fitting Larson's relation is defined as $R = \sqrt{A/\pi}$, where $A$ is the projected area of the cloud. The sample, available at Science Data Bank \citep{qi2026catalog}, spans $\sigma_v = 0.35$--$2.07$~km\,s$^{-1}$, $R = 0.26$--$8.66$~pc, and distance $D = 0.71$--$6.65$~kpc (1st--99th percentiles).

For each cloud, we apply an intensity threshold $T_\mathrm{th}$, keep only voxels with brightness temperature above $T_\mathrm{th}$, and recompute the velocity dispersion and effective radius at that threshold. The velocity dispersion is computed as the intensity-weighted second moment of the line-of-sight velocity,
\begin{equation}
\sigma_v = \sqrt{\sum T_i v_i^2 / \sum T_i - \left(\sum T_i v_i / \sum T_i\right)^2},
\end{equation}
where $T_i$ and $v_i$ are the brightness temperature and line-of-sight velocity of voxel $i$, and the sums run over all voxels above the applied threshold. We parameterize Larson's first law as
\begin{equation}
\sigma_v = C \cdot \left(\frac{R}{1\,\mathrm{pc}}\right)^\gamma,
\label{eq:larson}
\end{equation}
where $\gamma$ is the slope and $C$ is the normalization coefficient. We estimate the uncertainty in $\sigma_v$ for each cloud following the error-propagation formalism of \citet{furuya2014} (their Appendix~B), based on the per-pixel noise level; the median (90th-percentile) relative uncertainty is only 2.2\% (8.3\%). Because the per-pixel noise level is relatively uniform across the survey, we adopt a uniform noise level of 0.3~K; repeating the analysis with thresholds based on the per-pixel noise values gives consistent results.

Our fitting procedure consists of two steps. First, we scan $T_\mathrm{th}$ continuously until the sample size falls below $N = 50$ (see Figure~\ref{fig:gamma_curve}); at each threshold along this scan, we fit Equation~\eqref{eq:larson} to the corresponding sample via a linear regression in $\log R$--$\log \sigma_v$ space, obtaining the slope $\gamma(T_\mathrm{th})$ and the intercept $\log C(T_\mathrm{th})$. Second, to characterize how these parameters evolve with threshold, we fit the resulting sequences of $\gamma(T_\mathrm{th})$ and $C(T_\mathrm{th})$ with exponential saturation functions of $T_\mathrm{th}$, restricting this second fit to data points with $T_\mathrm{th} \leq 7$~K. Figure~\ref{fig:gamma_curve} shows all thresholds satisfying $N > 50$, which extend up to $T_\mathrm{th} = 9$~K; we adopt $T_\mathrm{th} = 7$~K as the upper limit of the fit because it lies close to the point where the sample size falls below $N = 100$ and the statistical uncertainties become too large for reliable fits. Statistical uncertainties come from the covariance matrix and are verified with 1000 bootstrap resamplings.
\section{Results} \label{sec:results}

\subsection{The \texorpdfstring{$\gamma(T_\mathrm{th})$}{gamma(T\_th)} Curve, the \texorpdfstring{$C(T_\mathrm{th})$}{C(T\_th)} Curve, and Fit Parameters}
\label{subsec:curve}

\begin{figure}[ht!]
\plotone{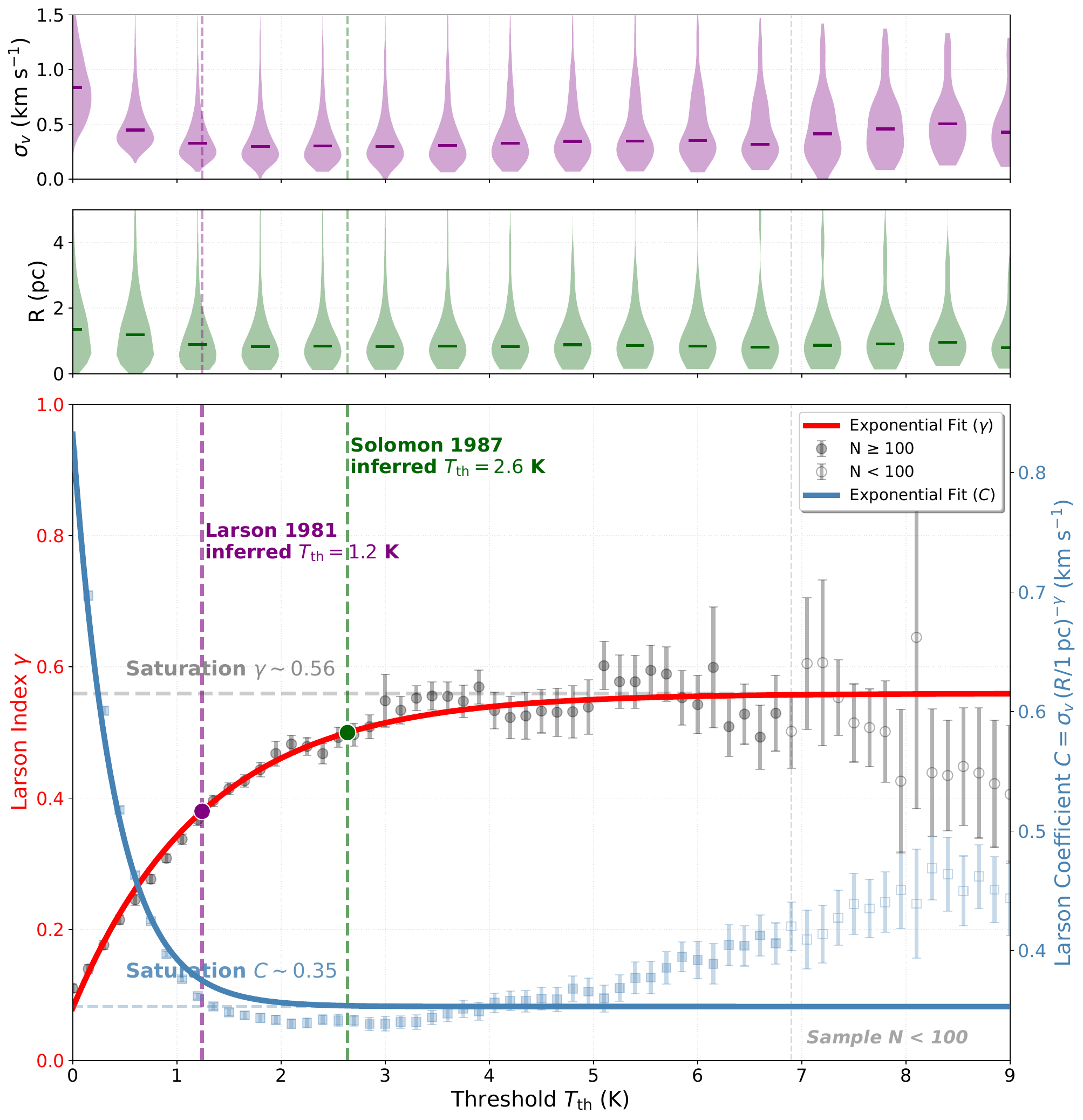}
\caption{Variation of the $\sigma_v$--$R$ index $\gamma$ (left axis, red) and the normalization coefficient $C$ (right axis, blue) with the intensity threshold $T_\mathrm{th}$ for MWISP $^{13}$CO molecular clouds. The top two panels show violin plots of the $\sigma_v$ and $R$ distributions at a subset of thresholds, with horizontal bars marking the median. In the bottom panel, filled markers indicate data points with $N \geq 100$ (with error bars); open markers have $N < 100$. The red and blue curves are the exponential saturation fits to $\gamma$ and $C$, respectively; the gray dashed line marks $\gamma_\mathrm{sat} \approx 0.56$, and the blue dashed line marks $C_\mathrm{sat} \approx 0.35$. The purple dashed line marks the threshold at which $\gamma = 0.38$ \citep{larson1981}, $T_\mathrm{th} \approx 1.2$~K, and the green dashed line that at which $\gamma = 0.5$ \citep{solomon1987}, $T_\mathrm{th} \approx 2.6$~K; both thresholds are inferred from the reported $\gamma$ values (Section~\ref{subsec:scatter}). The region to the right of the gray vertical line has $N < 100$.
\label{fig:gamma_curve}}
\end{figure}

Figure~\ref{fig:gamma_curve} shows the full variation of $\gamma$ with threshold. Starting near zero, $\gamma$ rises steeply up to $T_\mathrm{th} \sim 2.4$~K and then flattens. We fit a three-parameter exponential saturation model:
\begin{equation}
\gamma(T_\mathrm{th}) = \gamma_\mathrm{sat} - (\gamma_\mathrm{sat} - \gamma_0)\,\exp(-T_\mathrm{th}/T^*_\gamma),
\label{eq:gamma_fit}
\end{equation}
where $\gamma_\mathrm{sat}$ is the saturation index reached within the range of intensity thresholds probed by our data, $\gamma_0$ is the index at zero threshold, and $T^*_\gamma$ is the characteristic temperature at which $\gamma$ has completed $\sim$63\% of its rise. We note that $\gamma_\mathrm{sat}$ should not be interpreted as a true asymptotic value in the limit of arbitrarily fine spatial structure: the finite angular resolution of MWISP ($\sim52''$) sets a lower limit on the physical scales that can be resolved. The same linear regression simultaneously yields the normalization coefficient $C(T_\mathrm{th})$ (right axis of Figure~\ref{fig:gamma_curve}), which decreases monotonically with threshold and saturates. We fit $C(T_\mathrm{th})$ with an analogous exponential saturation form:
\begin{equation}
C(T_\mathrm{th}) = C_\mathrm{sat} + (C_0 - C_\mathrm{sat})\,\exp(-T_\mathrm{th}/T^*_C),
\label{eq:C_fit}
\end{equation}
where $C_\mathrm{sat}$ is the saturation normalization, $C_0$ is the normalization at zero threshold, and $T^*_C$ is the corresponding characteristic temperature. Best-fit parameters for both curves are listed in Table~\ref{tab:params}.

\begin{deluxetable}{lcc}
\tablecaption{Fit results for the $\gamma(T_{\rm th})$ and $C(T_{\rm th})$ curves obtained from the exponential saturation fits (Equations~\eqref{eq:gamma_fit} and \eqref{eq:C_fit}), using data points with $T_{\rm th} \leq 7.0$~K (see Figure~\ref{fig:gamma_curve}). Statistical uncertainties are 1$\sigma$, verified by bootstrap. \label{tab:params}}
\tablewidth{0pt}
\tablehead{
  \colhead{Parameter} & \colhead{Value} & \colhead{Description}
}
\startdata
$\gamma_0$            & $0.082 \pm 0.016$                      & index at zero threshold \\
$T^*_\gamma$        & $1.268 \pm 0.085$~K           & characteristic temperature ($\gamma$ curve) \\
$\gamma_\mathrm{sat}$ & $0.561 \pm 0.006$             & saturation index  \\
\hline
$C_0$                 & $0.832 \pm 0.015$~km\,s$^{-1}$           & normalization at zero threshold \\
$T^*_C$             & $0.401 \pm 0.023$~K           & characteristic temperature ($C$ curve) \\
$C_\mathrm{sat}$      & $0.353 \pm 0.003$~km\,s$^{-1}$ & saturation normalization \\
\enddata
\end{deluxetable}

We highlight three key findings from the $\gamma(T_\mathrm{th})$ curve. First, $\gamma_0 \approx 0.08$, indicating that the $\sigma_v$--$R$ relation essentially vanishes at zero threshold. At $T_{\rm th} = 0$~K, cloud boundaries extend into diffuse, low-brightness regions whose line profiles carry little coherent kinematic information, so including them adds to the measured size $R$ without contributing a corresponding increase in $\sigma_v$. Second, the purple line in Figure~\ref{fig:gamma_curve} marks the threshold position where $\gamma = 0.38$ ($T_\mathrm{th} \approx 1.2$~K), and the green line marks where $\gamma = 0.5$ ($T_\mathrm{th} \approx 2.6$~K). These two historically conflicting values may simply be different positions on the same curve, rather than independent physical conclusions. Third, the $\gamma(T_\mathrm{th})$ curve rises most steeply below the characteristic temperature $T^*_\gamma \approx 1.27$~K. In this range, a threshold difference of $\sim$0.3--0.6~K produces a large shift in $\gamma$. Since survey noise levels differ between studies, the same nominal threshold multiple maps to different actual brightness temperatures, placing measurements at different positions on the $\gamma(T_\mathrm{th})$ curve; this is a significant source of systematic scatter in reported $\gamma$ values that becomes transparent only when thresholds are expressed directly in brightness temperature.

The $C(T_\mathrm{th})$ curve complements these results. The coefficient $C$ drops rapidly at low thresholds and saturates at $C_\mathrm{sat} \approx 0.35$~km\,s$^{-1}$, with a characteristic temperature $T^*_C \approx 0.40$~K that is significantly lower than $T^*_\gamma$. The violin plots in the top two panels of Figure~\ref{fig:gamma_curve}, showing $\sigma_v$ and $R$, respectively, illustrate the underlying data evolution: the median $\sigma_v$ decreases systematically from $\sim$0.8~km\,s$^{-1}$ at low thresholds to $\sim$0.3~km\,s$^{-1}$ and then levels off, while the median $R$ decreases from $\sim$2.3~pc to $\sim$1.5~pc. This systematic evolution of the sample's velocity dispersion and size distributions provides the common observational basis for the rising $\gamma$ and falling $C$.

\begin{figure*}[ht!]
\plotone{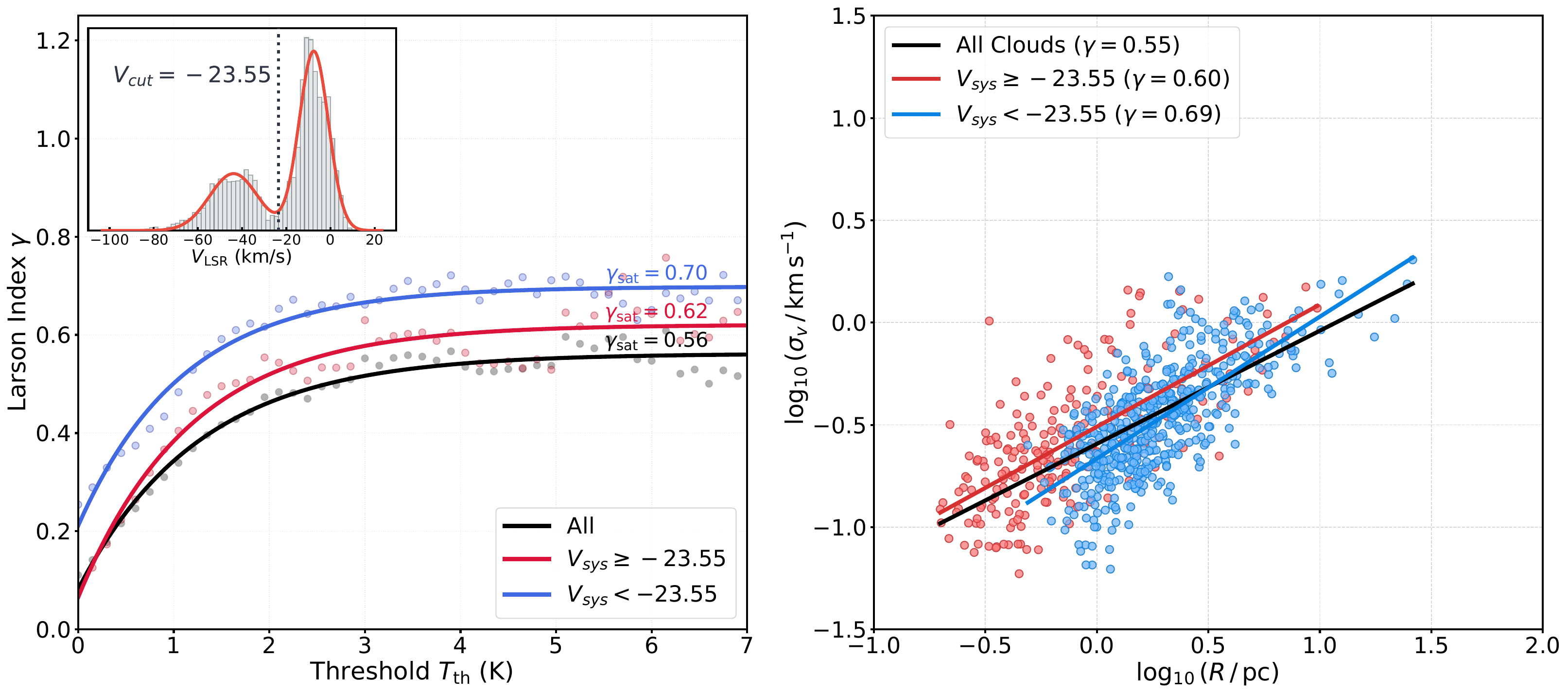}
\caption{Left: Variation of $\gamma$ with the intensity threshold $T_\mathrm{th}$ for the full sample (black) and the two velocity subgroups (red: $v_\mathrm{LSR} \geq -23.55$ km\,s$^{-1}$; blue: $v_\mathrm{LSR} < -23.55$ km\,s$^{-1}$). Exponential saturation fits are shown as solid lines; the saturation indices are $\gamma_\mathrm{sat} = 0.56$, $0.62$, and $0.70$ for the full sample and the two subgroups, respectively. Inset: line-of-sight velocity $v_\mathrm{LSR}$ distribution of the final sample; the red curve is the double-Gaussian fit, the blue and green dashed curves are the two components, and the dashed vertical line marks the crossing velocity $v_\mathrm{cut} = -23.55$ km\,s$^{-1}$. Right: $\log_{10}\sigma_v$--$\log_{10}R$ scatter plot at $T_\mathrm{th} = 3.3$~K for the three groups, with best-fit lines and corresponding $\gamma$ values in the legend.
\label{fig:groups}}
\end{figure*}

\subsection{Velocity Subgroups} \label{subsec:groups}

The full-sample $\gamma_\mathrm{sat}$ may be affected by mixing clouds at different distances and in different spiral-arm environments. To test how the converged index depends on sample composition, we split the sample by line-of-sight velocity. In the second quadrant, CO emission is confined to two velocity lanes, corresponding to the Local and Perseus arms \citep[see Figure~3 of][]{dame2001}, separated by a nearly empty gap \citep{digel1996}, and the two arms have been defined by $v_\mathrm{LSR}$ intervals \citep{brunt2002}. Our sample is also bimodal in $v_\mathrm{LSR}$ (inset of Figure~\ref{fig:groups}, left panel); a double-Gaussian fit gives a crossing point at $v_\mathrm{LSR} \approx -23.55$~km\,s$^{-1}$, close to the valley between the peaks, and the resulting grouping is broadly consistent with the spiral-arm model of \citet{reid2019} (Appendix~\ref{app:arms}). We accordingly split the sample into two velocity subgroups and repeat the analysis for each. Results are listed in Table~\ref{tab:groups}.

\begin{deluxetable}{lccccc}
\tablecaption{Fit parameters for the full sample and the velocity subgroups.\label{tab:groups}}
\tablewidth{0pt}
\tablehead{
  \colhead{Group} & \colhead{$N$} & \colhead{$\gamma_\mathrm{sat}$} & \colhead{$C_\mathrm{sat}$~(km\,s$^{-1}$)} & \colhead{$T^*_\gamma$~(K)} & \colhead{$\gamma$ at $T_\mathrm{th} = 1.2$~K}
}
\startdata
$v \geq -23.55$ km\,s$^{-1}$ & 1966 & $0.62 \pm 0.02$ & $0.435 \pm 0.003$ & $1.23 \pm 0.27$ & 0.42 \\
$v < -23.55$ km\,s$^{-1}$   & 2950 & $0.70 \pm 0.01$ & $0.315 \pm 0.002$ & $1.11 \pm 0.08$ & 0.53 \\
Full sample                 & 4916 & $0.56 \pm 0.01$ & $0.353 \pm 0.003$ & $1.27 \pm 0.09$ & 0.38 \\
\enddata
\end{deluxetable}

Both subgroups give $\gamma_\mathrm{sat}$ well above 0.5: $0.62\pm0.02$ for $v\geq-23.55$~km\,s$^{-1}$ and $0.70\pm0.01$ for $v<-23.55$~km\,s$^{-1}$. The difference, $\Delta\gamma_\mathrm{sat}=0.08$, is $\sim$3.6 times the statistical uncertainty; systematic errors are discussed in Section~\ref{sec:gamma_inf}. The normalizations differ more strongly, $C_\mathrm{sat}=0.435\pm0.003$ and $0.315\pm0.002$~km\,s$^{-1}$, respectively (Table~\ref{tab:groups}). Figure~\ref{fig:groups} (right panel) shows the $\log_{10}\sigma_v$--$\log_{10}R$ scatter plot at $T_\mathrm{th}=3.3$~K, where the higher-$C$ subgroup lies at smaller $R$. A joint fit therefore raises the small-$R$ end and lowers the large-$R$ end, flattening the slope to $\gamma=0.55$, compared with $0.60$ and $0.69$ for the two subgroups; the same effect places the full-sample $\gamma_\mathrm{sat}=0.56$ below both subgroup values. 

\section{Discussion} \label{sec:discussion}

\subsection{Origin of the Vanishing \texorpdfstring{$\sigma_v$--$R$}{sigma\_v--R} Relation at Low Threshold}
\label{subsec:lowthresh}

At zero or very low thresholds, the $\sigma_v$--$R$ relation vanishes, with $\gamma$ approaching zero. This may arise mainly from faint, relatively broad emission, which occurs both in the low-brightness outskirts of clouds and as the dominant emission component in some faint, small clouds. Because the velocity dispersion $\sigma_v$ is weighted by the squared velocity offset from the line centroid, even weak emission extending over a broad velocity range can make a substantial contribution to $\sigma_v$. This effect is particularly pronounced in small clouds, which generally have lower peak brightness temperatures and for which faint emission therefore constitutes a larger fraction of the total emission. Consequently, the velocity dispersions of small clouds are preferentially enhanced, reducing the contrast in $\sigma_v$ between small and large clouds and thus flattening the $\sigma_v$--$R$ relation. As the threshold increases, this faint emission is progressively excluded, revealing a stronger dependence of velocity dispersion on spatial scale, with $\gamma$ increasing accordingly.

The faint and broad profiles in small clouds and in the outskirts of big ones are not necessarily genuine features. Gaussian decomposition may introduce broad components by fitting non-Gaussian line wings or noise fluctuations, while other cloud-identification algorithms may similarly incorporate noise into apparently coherent, low-brightness structures. Further investigation using dedicated simulations and deeper observations is therefore needed to establish what fraction of these features are real. To examine whether the increasing $\gamma$ depends on the inclusion of these small clouds with faint, relatively broad profiles, we repeat the analysis using only clouds that remain identified at $T_\mathrm{th} = 3.0$~K. The resulting $\gamma$ also increases with the threshold and approaches a saturation value of $\gamma_\mathrm{sat} \approx 0.56$. This shows that the rise and subsequent saturation of $\gamma$ are not driven solely by these small clouds, but persist even when they are excluded.

\subsection{The Meaning of \texorpdfstring{$T^*$}{T*} and Systematic Threshold Effects on \texorpdfstring{$\gamma$}{gamma}} 
\label{subsec:scatter}

The $\gamma(T_\mathrm{th})$ curve rises most steeply below $T^*_\gamma = 1.27 \pm 0.09$~K, where $\gamma$ is highly sensitive to the choice of threshold; for the MWISP noise level of $\sim$0.3~K, standard thresholds of 2--3 times this value ($T_\mathrm{th} \approx 0.6$--$0.9$~K) fall on this flank. This steep transition offers a natural explanation for the historical scatter in reported $\gamma$: studies whose effective thresholds fall on different parts of this flank, whether due to differences in noise level, threshold multiplier, or boundary definition, will recover systematically different $\gamma$ even for similar clouds, unless thresholds are matched on a common brightness-temperature scale \citep{heyerdame2015}.

When the two classic indices are placed on our curve (Figure~\ref{fig:gamma_curve}), $\gamma \approx 0.38$ \citep{larson1981} corresponds to $T_\mathrm{th} \approx 1.2$~K and $\gamma \approx 0.5$ \citep{solomon1987} to $T_\mathrm{th} \approx 2.6$~K. Larson's compilation applied no uniform threshold: in 37 of the 38 references we checked, small clouds were mostly bounded by half-maximum contours, and large clouds mostly by the outermost detectable contour. The best-determined $^{13}$CO boundary temperatures for small clouds are $T_A^* \approx 1.3$--1.8~K for five dark globules \citep{martin1978} and above 2~K for NGC~2023--24 \citep{milman1975}. Three large clouds were bounded in $^{12}$CO, at about 2--2.5~K for L1641 and Per~OB2 \citep{kutner1977, sargent1979}, corresponding to an even lower threshold in $^{13}$CO because $^{12}$CO is usually stronger. These effective thresholds are roughly consistent with the 1.2~K from our curve. The match is only approximate: temperature scales differ among studies, and $\sigma$ was not always measured within the boundary defining the size. We therefore offer this as a plausible explanation rather than a proof. \citet{solomon1987} identified clouds as closed $^{12}$CO isointensity surfaces at $T_\mathrm{min} = 4$~K by default, raised to 5--7~K in confused regions and lowered to 3~K at $l > 54^\circ$, while $\sigma_v$ and the size were computed from all emission above 1~K inside the box circumscribing each surface. Their thresholds are defined on the $T_R^*$ scale, whereas MWISP uses the main-beam temperature $T_\mathrm{mb}$. Since the FCRAO 14~m and PMO 13.7~m telescopes share the same antenna design, we adopt $\eta_\mathrm{fss} = 0.7$ and $\eta_\mathrm{mb} = 0.45$ \citep{sun2021}, giving $T_\mathrm{mb} = (\eta_\mathrm{fss}/\eta_\mathrm{mb})\,T_R^* \approx 1.56\,T_R^*$. The main remaining uncertainty is the $^{13}$CO/$^{12}$CO brightness temperature ratio, which in MWISP data is concentrated at $\sim$0.15--0.45 \citep{wang2023}. The 4~K threshold therefore corresponds to $\sim$0.9--2.8~K in $^{13}$CO, and the full 3--7~K range to $\sim$0.7--4.9~K. The $T_\mathrm{th}\simeq 2.6$~K from our curve in Figure~\ref{fig:gamma_curve}, corresponding to $\gamma\simeq0.5$, lies within the converted range for the default 4~K threshold. The $\gamma \approx 0.38$ versus $0.5$ difference may thus largely reflect differences in detection depth rather than distinct physical conditions.
The normalization $C(T_\mathrm{th})$ behaves differently: $T^*_C \approx 0.40$~K, well below $T^*_\gamma \approx 1.27$~K, so $C$ saturates well before $\gamma$ does. Both parameters share one regression, so the within-cloud stripping seen in the fixed-subsample test (Section~\ref{subsec:lowthresh}) can plausibly account for $C$'s decline alongside $\gamma$'s rise. Whether the gap between $T^*_C$ and $T^*_\gamma$ reflects an additional physical effect, for instance a contribution from changing sample composition at low threshold, or simply reflects how $C$ and $\gamma$ respond differently to the same underlying stripping process, is not clear from the present data.

\subsection{The Meaning of \texorpdfstring{$\gamma_{\rm sat}$}{gamma\_sat} and \texorpdfstring{$C_{\rm sat}$}{C\_sat}, and Factors Affecting Their Values}
\label{sec:gamma_inf}

The saturation index $\gamma_{\rm sat}$ denotes the exponent of the relation between $\sigma_v$ and $R$ at high thresholds, where dense cores dominate, and reflects the dynamical state of these clouds' dense internal structure. In the framework of \citet{heyer2009}, a virialized cloud satisfies \begin{equation} \gamma = 0.5 + 0.5 \frac{d\log\Sigma}{d\log R}, \end{equation} where $\gamma = 0.5$ corresponds to a scale-independent surface density. For the full sample, $\gamma_{\rm sat} = 0.56$, slightly above 0.5, indicating a mild positive correlation between surface density and scale. However, \citet{bp2006} noted that the $\sigma_v$--$R$ relation requires only that kinetic and gravitational energies be approximately comparable, not strict virial equilibrium; in supersonic turbulence, mass and momentum are continuously exchanged, making strict virial equilibrium difficult to sustain. A $\gamma_{\rm sat}$ slightly above 0.5 therefore reflects a tendency toward energy equipartition rather than strict virial equilibrium.

As noted in Section~\ref{sec:data}, the saturation fit is restricted to $T_{\rm th} \leq 7$~K; varying this limit between 6~K and 9~K changes $\gamma_{\rm sat}$ only slightly, within $\sim$0.56--0.58. Kinematic distance errors, which enter through $R$, could affect both the shape and the values of the $\gamma(T_{\rm th})$ curve (Appendix~\ref{subsec:app_distance}). The shape is robust: replacing $R$ with the distance-independent angular size $\theta$ reproduces the same rising-and-saturating trend (Figure~\ref{fig:A4}). The values, however, are biased downward because random errors in $R$ flatten the fitted slope: in a Monte Carlo test, perturbing cloud velocities lowers the whole curve (Figure~\ref{fig:A2}), and a bivariate correlated errors and intrinsic scatter (BCES) refit estimator that accounts for these errors steepens the slope (Figure~\ref{fig:A3}). The reported $\gamma(T_{\rm th})$ values, including $\gamma_{\rm sat}$, are therefore conservative lower bounds.

The elevated $\gamma$ values reported by \citet{miville2017} ($\approx 0.63$) and \citet{sun2024} ($\approx 0.66$; see Section~\ref{sec:intro}) may partly reflect error-corrected fitting, as noted above, and a bias shared by both studies: \citet{miville2017} merged multiple velocity components into single clouds, and the clouds of \citet{sun2024}, identified with density-based spatial clustering of applications with noise (DBSCAN), may be subject to the same blending (Section~\ref{subsec:12co}). Blending preferentially affects large clouds, inflating $\sigma_v$ and steepening the slope.

The value of $\gamma_{\rm sat}$ depends on sample composition. When the sample is divided by line-of-sight velocity, the median distance within each subgroup remains nearly constant across thresholds, whereas the two subgroups reach $\gamma_{\rm sat}$ values ($0.62 \pm 0.02$ and $0.70 \pm 0.01$) well above the full-sample value of 0.56, and differ even more strongly in normalization ($\Delta C_\mathrm{sat}\approx0.12$~km\,s$^{-1}$; Table~\ref{tab:groups}). This raises the question of whether dense clouds within a single environment deviate from virial equilibrium more than the full sample suggests, or whether the higher subgroup values instead trace some other systematic effect. Since both subgroups come from the same regression, which is affected by kinematic distance errors that can lower $\gamma$ by 0.08--0.14 (Appendix~\ref{subsec:app_distance}) and need not do so equally in the two subgroups, the difference in slope ($\Delta\gamma_\mathrm{sat}=0.08$) may be observational in origin (distance errors, different size ranges caused by sample selection effects, or blending of unresolved clouds) or reflect physical differences between the subgroups. To test whether the lack of small clouds in the $v<-23.55$~km\,s$^{-1}$ subgroup explains the difference, we take $T_{\rm th}=3.3$~K (Figure~\ref{fig:groups}, right panel) as an example, where $\gamma$ is already close to saturation, and apply the corresponding lower size limits, $R>1.0$~pc and $R>1.5$~pc, to the other subgroup. This lowers its $\gamma$ from 0.60 to 0.58 and 0.51, respectively, instead of raising it toward 0.69, so the lack of small clouds is unlikely to explain the difference. Physically, \citet{xie2025} proposed that, if clouds deviate from energy equipartition, large-scale shear can raise $\gamma$ above 0.5, so a difference in shear strength could contribute; the detailed mechanism is left for future work.

The same regression also yields a saturation normalization of $C_{\rm sat} \approx 0.35$~km\,s$^{-1}$. Under the approximation $\gamma \approx 0.5$, the framework proposed by \citet{heyer2009} for clouds in virial equilibrium gives $\sigma_v = (\pi G \Sigma / 5)^{0.5}R^{0.5}$. Comparing this with $\sigma_v = C \cdot (R/1\,{\rm pc})^\gamma$ gives $C = (\pi G \Sigma / 5)^{0.5}\,(1\,{\rm pc})^{0.5}$, so that the median surface density can be inferred from the normalization coefficient as
\begin{equation}
\Sigma = \frac{5 C_{\rm sat}^2}{\pi G\,(1\,{\rm pc})} \approx 46~M_\odot\,{\rm pc}^{-2}.
\end{equation}
This value is consistent with the median surface density of $42~M_\odot\,{\rm pc}^{-2}$ independently derived by \citet{heyer2009} for a comparable sample from $^{13}$CO column densities assuming local thermodynamic equilibrium.

\begin{figure*}[ht!]
\centering
\includegraphics[width=0.85\textwidth]{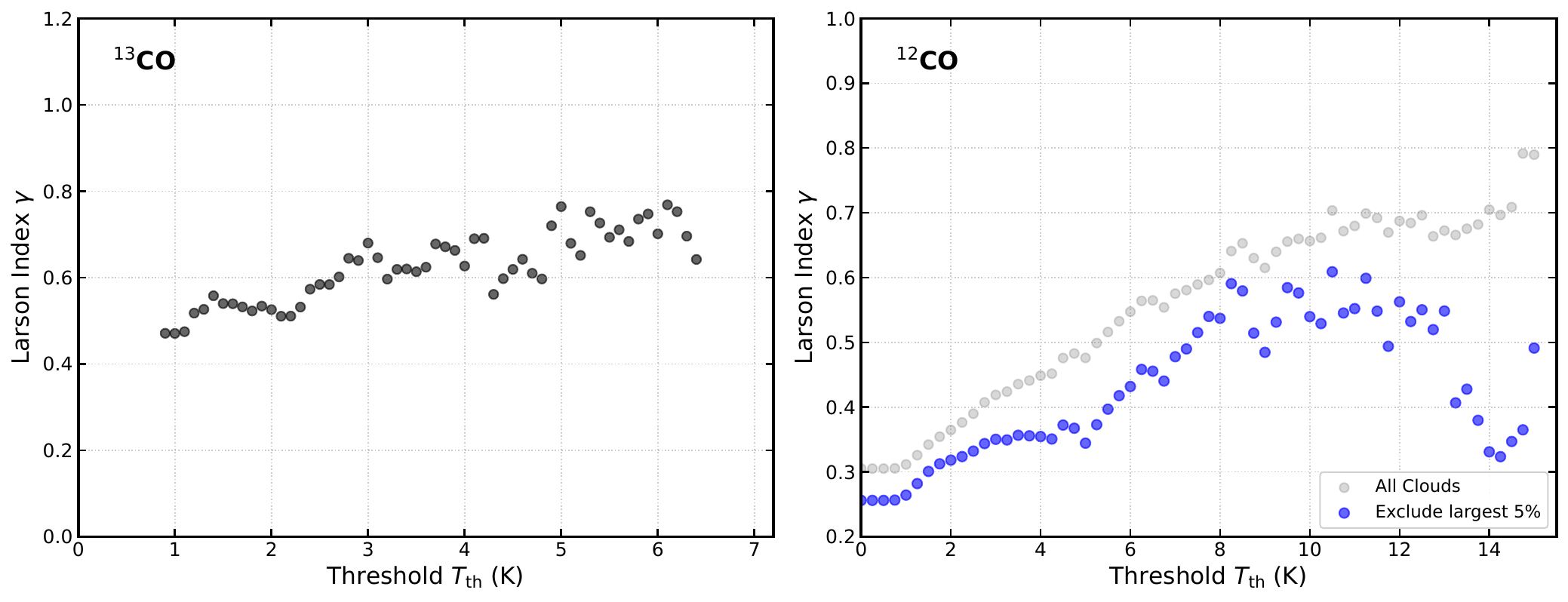}
\caption{Variation of the $\sigma_v$--$R$ index $\gamma$ with the intensity threshold $T_\mathrm{th}$ for two DBSCAN-identified samples \citep{yan2021, yuan2022}. \emph{Left:} MWISP $^{13}$CO sample (same tracer as in Figure~\ref{fig:gamma_curve}). \emph{Right:} MWISP $^{12}$CO sample; gray circles show the full cloud sample, and blue circles show the sample after excluding the largest 5\% of clouds by projected size. Both panels show $\gamma$ rising monotonically without clear saturation, unlike the saturating $^{13}$CO/GDCluster result in Figure~\ref{fig:gamma_curve}.
\label{fig:gamma_dbscan}}
\end{figure*}

\subsection{Comparison with DBSCAN-Identified Samples} \label{subsec:12co}

To test whether the monotonic increase in $\gamma(T_\mathrm{th})$ is an artifact of the cloud-finding algorithm, the tracer, or both, we compare our fiducial $^{13}$CO/GDCluster result (Figure~\ref{fig:gamma_curve}) with two DBSCAN-identified samples \citep{yuan2022, yan2021}: $^{13}$CO (Figure~\ref{fig:gamma_dbscan}, left) and $^{12}$CO (Figure~\ref{fig:gamma_dbscan}, right). Both DBSCAN samples show $\gamma$ rising monotonically without clear saturation, even for the $^{13}$CO sample, which uses the same tracer and the same MWISP raw data as our fiducial sample. This indicates that the saturating behavior may arise primarily from the algorithm rather than the tracer. Unlike GDCluster, DBSCAN can merge different velocity components into a single cloud; these blended components persist at high thresholds, broadening $\sigma_v$ and preventing $\gamma$ from converging.

Removing the largest 5\% of clouds from the $^{12}$CO sample (blue points, Figure~\ref{fig:gamma_dbscan}, right) reduces $\gamma$ at all thresholds; the same test on the $^{13}$CO/GDCluster sample also lowers $\gamma_\mathrm{sat}$, showing this size-bias effect is not limited to the blending of multiple velocity components into single clouds. Whether the large-cloud contribution is dominated by shear or unresolved multiple components cannot be disentangled with the present data and is left for future work.

\section{Conclusions} \label{sec:conclusions}
Using 4916 $^{13}$CO molecular clouds identified with GDCluster \citep{liu2025} in the MWISP survey, we measure the power-law index $\gamma$ of the velocity dispersion--size relation, $\sigma_v = C\,(R/1\,\mathrm{pc})^\gamma$, as a continuous function of the intensity threshold $T_{\rm th}$. Our main conclusions are as follows.
\begin{enumerate}
\item The index $\gamma$ rises monotonically with threshold, from $\gamma_0 \approx 0.08$ at zero threshold to $\gamma_{\rm sat} \approx 0.56$ at saturation, with the steepest increase below $T^*_\gamma \approx 1.27$~K. Because kinematic distance errors bias $\gamma$ downward, the reported $\gamma(T_{\rm th})$ values, including $\gamma_{\rm sat}$, are conservative lower bounds. The normalization coefficient $C(T_{\rm th})$ correspondingly saturates at $C_{\rm sat} \approx 0.35$~km\,s$^{-1}$, implying a surface density of $\sim$46~$M_\odot\,{\rm pc}^{-2}$ under a virial approximation. The two historically debated values of $\gamma$ may largely reflect different positions on this curve: the $\gamma \approx 0.38$ of \citet{larson1981} corresponds to $T_{\rm th} \approx 1.2$~K on our curve, while the $\gamma \approx 0.5$ of \citet{solomon1987} corresponds to a higher effective threshold near saturation, $T_{\rm th} \approx 2.6$~K. This steep rise of $\gamma$ with threshold is a previously uncharacterized source of systematic scatter in reported $\gamma$ values, and should be considered in future studies of Larson-type relations and in assessing the roles of turbulence and gravity in shaping molecular cloud structure.
\item The full-sample $\gamma_{\rm sat}$ depends on sample composition. Splitting the sample by line-of-sight velocity (the Local and Perseus arms) yields higher $\gamma_{\rm sat}$ ($0.62$ and $0.70$) and different normalizations ($C_{\rm sat}=0.435$ and $0.315$~km\,s$^{-1}$), but the current analysis cannot determine whether this is an observational effect or has a deeper physical origin. 
\item Independent $^{13}$CO and $^{12}$CO samples identified with DBSCAN reproduce the same monotonically rising trend in $\gamma$, confirming that this effect is robust across tracers and cloud-finding methods; the lack of saturation is likely attributable to the cloud-finding algorithm.
\end{enumerate}

\begin{acknowledgments}
We thank the referee for very helpful comments. This work is supported by the National Key R\&D Program of China under grant 2023YFA1608000 and the National SKA Program of China under grant 2025SKA0140100.
\end{acknowledgments}

\facilities{PMO:DLH}

\apptablenumbers
\appendix
\section{Distance Reliability of the Sample}
\label{app:distance}

\renewcommand{\topfraction}{0.9}
\renewcommand{\bottomfraction}{0.9}
\renewcommand{\textfraction}{0.05}
\renewcommand{\floatpagefraction}{0.5}
\setcounter{topnumber}{4}
\setcounter{bottomnumber}{4}
\setcounter{totalnumber}{6}

\subsection{Sensitivity of the Subgroup Comparison to the \texorpdfstring{$|v_{\rm LSR}|$}{|v\_LSR|} Cut}
\label{subsec:app_vcut}

\begin{figure}[htbp]
\centering
\includegraphics[width=0.8\columnwidth]{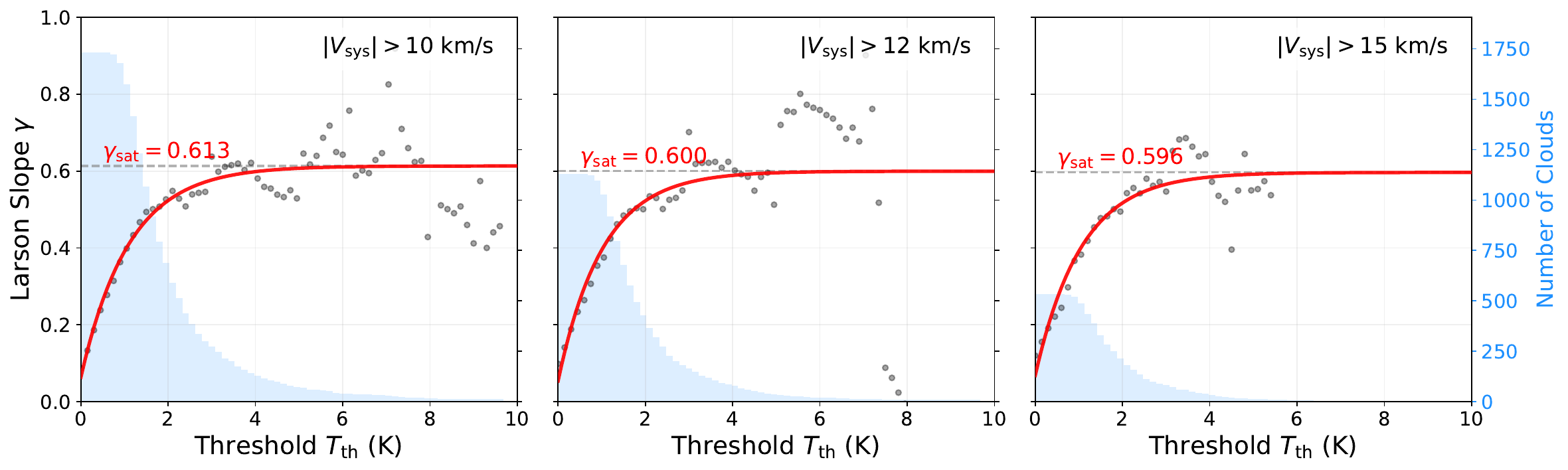}
\caption{Sensitivity of $\gamma_{\rm sat}$, for the $v \geq -23.55$~km\,s$^{-1}$ subgroup, to the $|v_{\rm LSR}|$ cut applied in Section~\ref{sec:data}, raised from 10 to 12 and 15~km\,s$^{-1}$.}
\label{fig:A1}
\end{figure}

Since all $v < -23.55$~km\,s$^{-1}$ clouds have $|v_{\rm LSR}| > 15$~km\,s$^{-1}$, raising the $|v_{\rm LSR}|$ cut to 12 and 15~km\,s$^{-1}$ removes clouds only from the $v \geq -23.55$~km\,s$^{-1}$ subgroup. Refitting this subgroup alone therefore isolates the effect of the cut from the mixing between the two velocity subgroups: $\gamma_{\rm sat}$ changes only from 0.613 to 0.600 and 0.596 (Figure~\ref{fig:A1}), within the fit uncertainty of $\pm 0.02$, while the sample size drops rapidly.

\FloatBarrier
\subsection{Robustness of \texorpdfstring{$\gamma_{\rm sat}$}{gamma\_sat} to Kinematic Distance Errors}
\label{subsec:app_distance}

\begin{figure}[htbp]
\centering
\includegraphics[width=0.85\columnwidth]{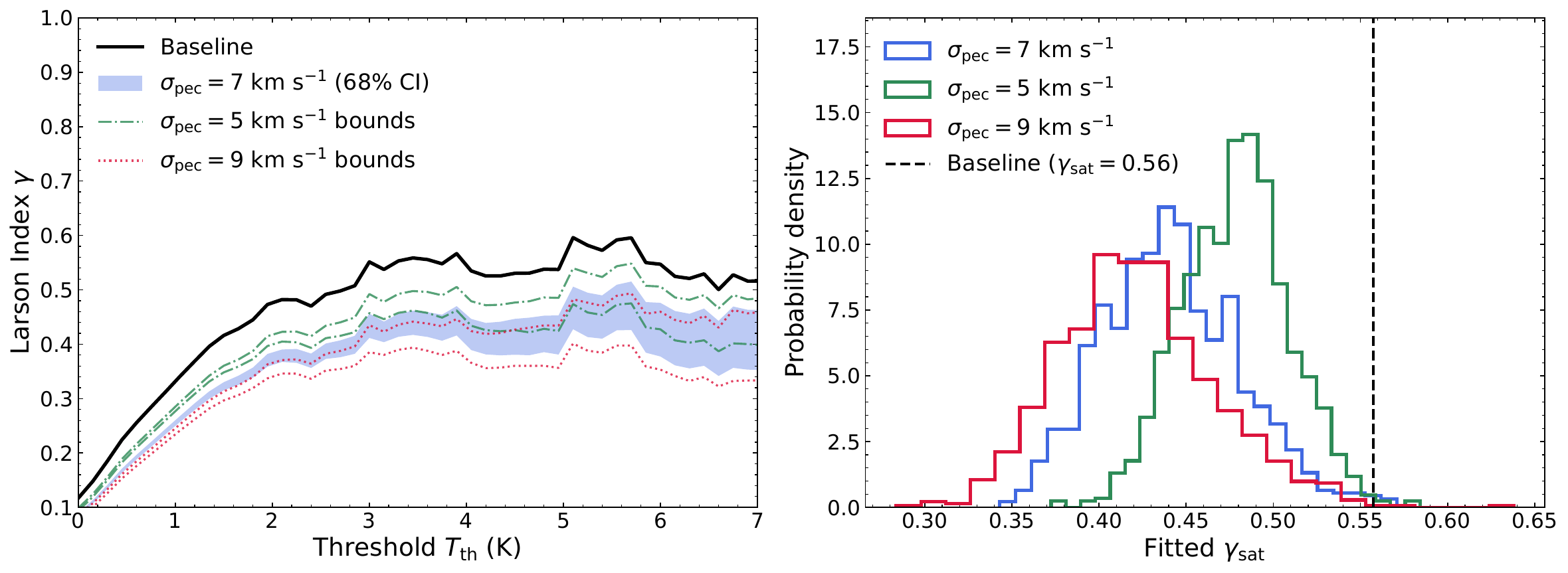}
\caption{Monte Carlo test of the effect of kinematic distance errors on $\gamma_{\rm sat}$, for the full sample. \emph{Left:} $\gamma(T_{\rm th})$ curves recovered after perturbing each cloud's systemic velocity by a peculiar-velocity term drawn from $\mathcal{N}(0, \sigma_{\rm pec})$ and recomputing distances, $R$, and the threshold scan; the solid line and shaded band show the median and 68\% confidence interval for $\sigma_{\rm pec} = 7$~km\,s$^{-1}$ (1000 resamplings), and the dashed lines mark the corresponding bounds for $\sigma_{\rm pec} = 5$ and 9~km\,s$^{-1}$. \emph{Right:} distribution of the fitted $\gamma_{\rm sat}$ across resamplings for each $\sigma_{\rm pec}$, compared to the baseline (unperturbed) value $\gamma_{\rm sat} = 0.56$ (vertical line).}
\label{fig:A2}
\end{figure}

\begin{figure}[htbp]
\centering
\includegraphics[width=0.45\columnwidth]{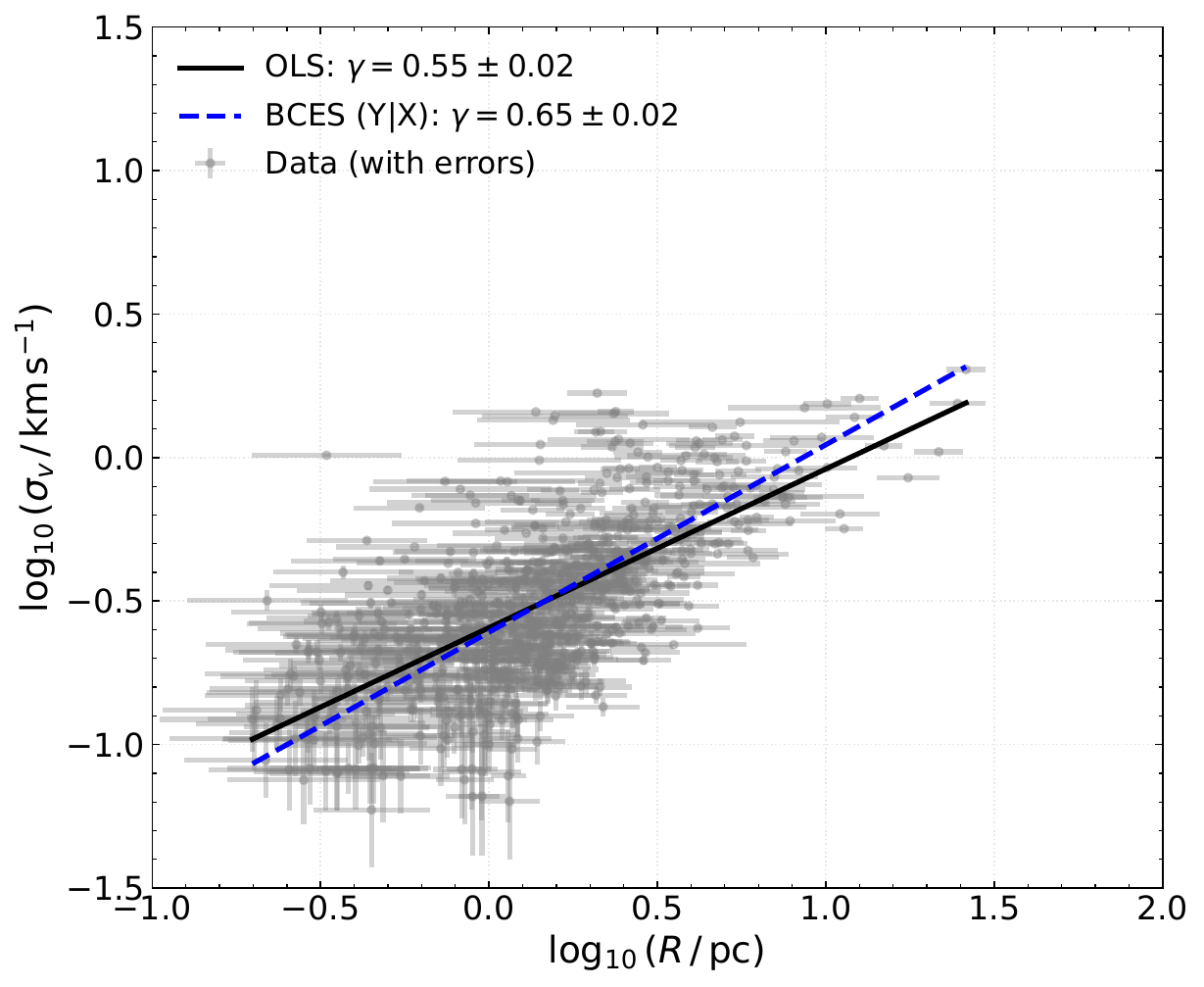}
\caption{Unweighted ordinary least-squares fit (slope $= 0.553 \pm 0.022$) and BCES(Y$|$X) fit \citep[][slope $= 0.654 \pm 0.024$]{akritas1996} to the $\sigma_v$--$R$ relation at $T_{\rm th} = 3.3$~K ($|v_{\rm LSR}| > 10$~km\,s$^{-1}$ subsample, $N = 732$), illustrating the correction for regression dilution.}
\label{fig:A3}
\end{figure}

A Monte Carlo test perturbs each cloud's systemic velocity by $\mathcal{N}(0, \sigma_{\rm pec})$, with $\sigma_{\rm pec} = 5$, 7, and 9~km\,s$^{-1}$ (informed by \citealt{reid2019, hunter2024}; 1000 resamplings each), and recomputes distances, $R$, and $\gamma(T_{\rm th})$. The recovered $\gamma_{\rm sat}$ decreases monotonically from 0.56 (unperturbed) to 0.48, 0.45, and 0.42 (Figure~\ref{fig:A2}), showing that kinematic distance errors bias $\gamma$ downward through regression dilution. This test targets random errors; systematic errors from non-axisymmetric structure are smaller within spiral arms than between them \citep{hunter2024}, so our arm-tracing sample is less affected. Correcting for this bias directly in the observed sample gives the same picture: refitting $\sigma_v$--$R$ at $T_{\rm th} = 3.3$~K with the BCES(Y$|$X) estimator \citep{akritas1996} raises the slope from $0.553 \pm 0.022$ to $0.654 \pm 0.024$ (Figure~\ref{fig:A3}). Since this bias enters at every threshold, the entire $\gamma(T_{\rm th})$ curve should be regarded as a conservative lower bound.

\begin{figure}[htbp]
\centering
\includegraphics[width=0.45\columnwidth]{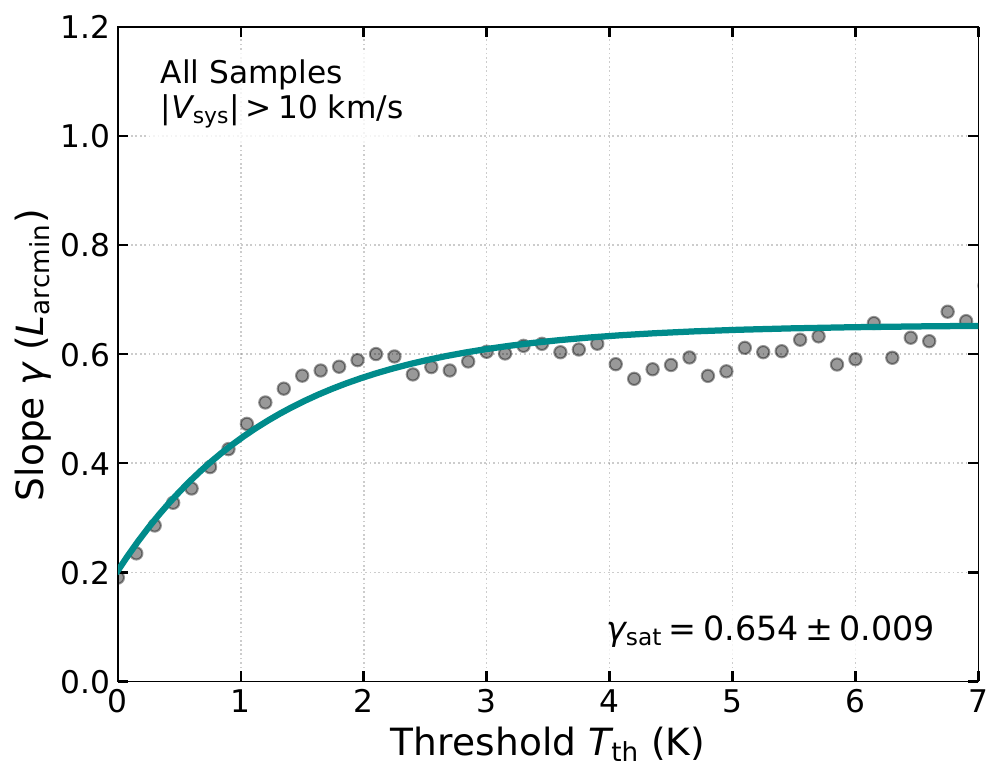}
\caption{Threshold scan repeated using each cloud's angular size $\theta$ in place of the physical radius $R$. Gray points show the fitted slope of the $\sigma_v$--$\theta$ relation at each threshold $T_{\rm th}$; the solid line is the exponential saturation fit, which saturates at $0.654 \pm 0.009$.}
\label{fig:A4}
\end{figure}
Finally, repeating the threshold scan with the distance-independent angular size $\theta$ in place of $R$ reproduces the same rising-and-saturating shape (Figure~\ref{fig:A4}). The saturation value, $0.654 \pm 0.009$, is not directly comparable to $\gamma_{\rm sat}$ because $\theta$ is not a physical size; this check shows only that the shape is not an artifact of distance errors.

\FloatBarrier
\section{Spiral-Arm Assignment of the Velocity Subgroups}
\label{app:arms}

\begin{figure}[ht!]
\centering
\includegraphics[width=0.52\columnwidth]{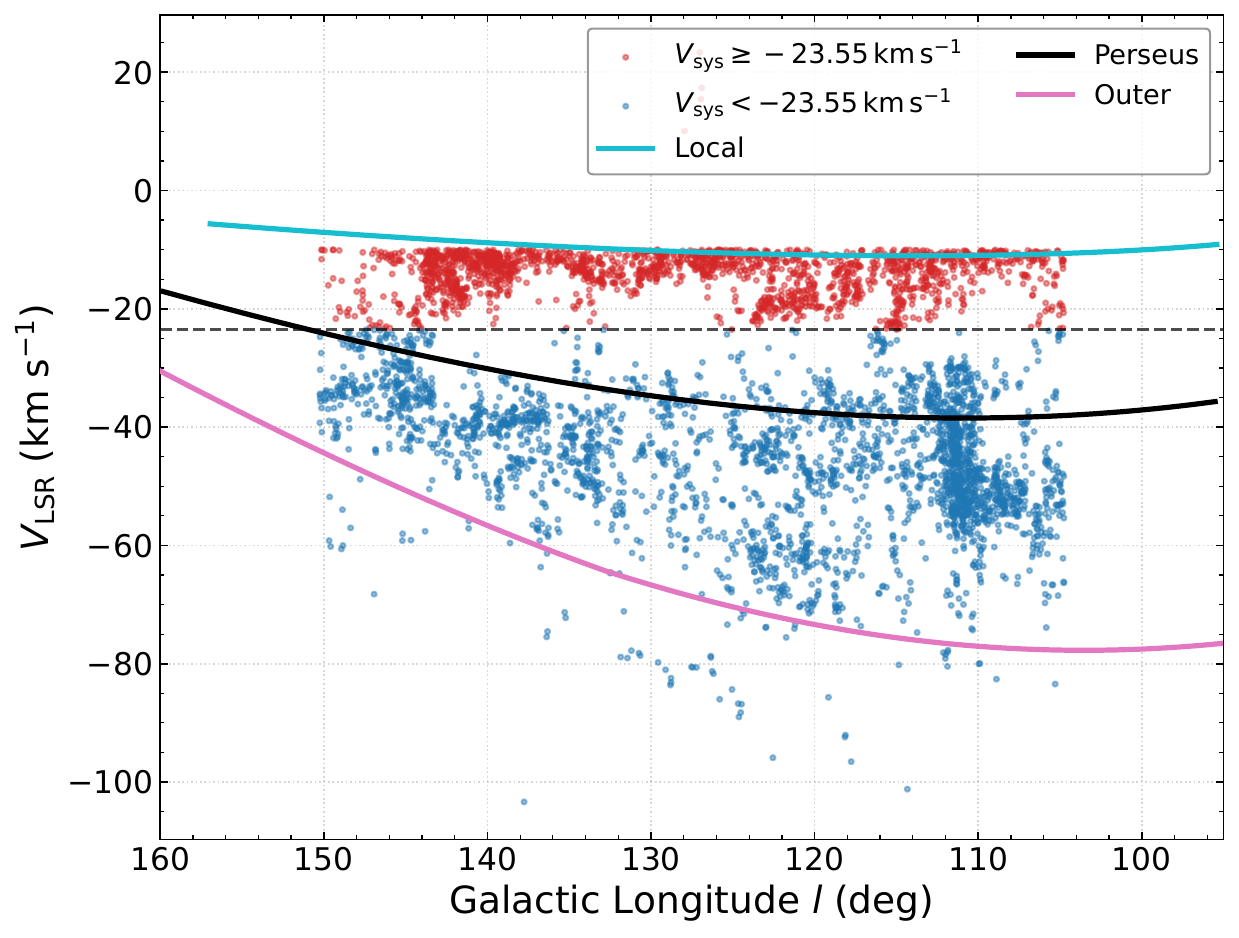}
\caption{Distribution of the sample in the $(l, v_{\rm LSR})$ plane, colored by velocity subgroup, overplotted with the spiral-arm model of \citet[][Table~2]{reid2019}. The $v \geq -23.55$~km\,s$^{-1}$ subgroup (red) falls predominantly along the Local arm; the $v < -23.55$~km\,s$^{-1}$ subgroup (blue) falls along the Perseus arm and extends toward the Outer arm.}
\label{fig:B1}
\end{figure}

Figure~\ref{fig:B1} compares the velocity subgroups against the spiral-arm model of \citet[][Table~2]{reid2019}. A constant velocity cut, rather than a point-by-point assignment along the model curve, is adopted as a practical approximation: in this longitude range the clouds' non-circular motions are comparable to the velocity width of the arms in the $(l, v_{\rm LSR})$ plane, so a point-by-point assignment would not be accurate either.

\clearpage
\bibliography{manuscript_aastex701}{}
\bibliographystyle{aasjournalv7}

\end{document}